\documentclass[fleqn,usenatbib,useAMS]{mnras}

\usepackage{graphicx}
\usepackage{amsmath}
\usepackage{amssymb}
\usepackage{multicol}      
\usepackage{bm}	
\usepackage{pdflscape}
\usepackage{threeparttable}
\usepackage{longtable}
\usepackage{multirow}

\usepackage[T1]{fontenc}

\usepackage{newtxtext,newtxmath}

\title[Evidence of Deep Mixing in IRS 7]{\textit{Evidence of Deep Mixing in IRS 7, a Cool Massive Supergiant Member of the Galactic Nuclear Star Cluster}}

\author[Rafael Guerço et al.]{
Rafael Guer\c{c}o,$^{1,2}$\thanks{Contact e-mail: \href{mailto:guercorafael@gmail.com}{guercorafael@gmail.com}}
Verne V. Smith,$^{3}$
Katia Cunha,$^{2,4}$
Sylvia Ekstr\"om,$^{5}$
Carlos Abia,$^{6}$
Bertrand Plez,$^{7}$
\newauthor
Georges Meynet,$^{5}$
Solange V. Ramirez,$^{8}$
Nikos Prantzos,$^{9}$
Kris Sellgren,$^{10}$
Cristian R. Hayes$^{11}$ and
\newauthor
Steven R. Majewski$^{12}$
\\
$^{1}$Centro Educacional Chambarelli, Quintino Bocaiuva, Rio de Janeiro, RJ, Brazil\\
$^{2}$Observat\'orio Nacional, S\~ao Crist\'ov\~ao, Rio de Janeiro, RJ, Brazil\\
$^{3}$NOIRLab, Tucson, AZ 85719 USA\\
$^{4}$University of Arizona, Tucson, AZ 85719, USA\\
$^{5}$Universit\'e de Geneve, Chemin Pegasi 51, CH-1290 Versoix, Switzerland\\
$^{6}$Dpto. F\'\i sica Te\'orica y del Cosmos, 18071 Universidad de Granada, Granada, Spain\\
$^{7}$LUPM, Univ Montpellier, CNRS, Montpellier, France\\
$^{8}$Carnegie Observatories, 813 Santa Barbara Street, Pasadena, CA 91101, USA\\
$^{9}$Institut d'Astrophysique de Paris, UMR7095 CNRS, Sorbonne Université, 98bis Bd. Arago, F-75104 Paris, France\\
$^{10}$The Ohio State University, 140 West 18th Avenue, Columbus, OH 43210, USA\\
$^{11}$NRC Herzberg Institute of Astrophysics, Victoria, BC, V9E 2E7, Canada\\
$^{12}$University of Virginia, Charlottesville, VA 22904-4325, USA
}

\date{Accepted 2022 August 20. Received 2022 August 19; in original form 2022 July 6}

\pubyear{2022}

\begin{document}
\label{firstpage}
\pagerange{\pageref{firstpage}--\pageref{lastpage}}
\maketitle

\begin{abstract}
The center of the Milky Way contains stellar populations spanning a range in age and metallicity, with a recent star formation burst producing young and massive stars. 
Chemical abundances in the most luminous stellar member of the Nuclear Star Cluster (NSC), IRS 7, are presented for $^{19}$F, $^{12}$C, $^{13}$C, $^{14}$N, $^{16}$O, $^{17}$O, and Fe from an LTE analysis based on spherical modeling and radiative transfer with a 25M$_{\odot}$ model atmosphere, whose chemistry was tailored to the derived photospheric abundances. We find IRS 7 to be depleted heavily in both $^{12}$C ($\sim$-0.8 dex) and $^{16}$O ($\sim$-0.4 dex), while exhibiting an extremely enhanced $^{14}$N abundance ($\sim$+1.1 dex), which are isotopic signatures of the deep mixing of CNO-cycled material to the stellar surface. 
The $^{19}$F abundance is also heavily depleted by $\sim$1 dex relative to the baseline fluorine of the Nuclear Star Cluster, providing evidence that fluorine along with carbon constrain the nature of the deep mixing in this very luminous supergiant. 
The abundances of the minor isotopes $^{13}$C and $^{17}$O are also derived, with ratios of $^{12}$C/$^{13}$C$\sim$5.3 and $^{16}$O/$^{17}$O$\sim$525. 
The derived abundances for IRS 7, in conjunction with previous abundance results for massive stars in the NSC, are compared with rotating and non-rotating models of massive stars and it is found that the IRS 7 abundances overall follow the behavior predicted by stellar models. The depleted fluorine abundance in IRS 7 illustrates, for the first time, the potential of using the $^{19}$F abundance as a mixing probe in luminous red giants.
\end{abstract}


\section{Introduction}

The nuclear star cluster (NSC) located at the center of the Milky Way is a complex, massive agglomeration of stars (M$\sim$2.5$\times$ 10$^{7}$M$_{\odot}$, e.g., Sormani et al. \citeyear{somarni2020}) that surrounds the supermassive black hole Sgr A$^{*}$.  The NSC effective radius is 5 pc (Gallego-Cano et al. \citeyear{galego2020}), corresponding to $\sim$125" on the sky at a distance of 8.178 kpc (Gravity Collaboration et al. \citeyear{gravitycollaboration2019}), resulting in heavy observed stellar crowding in a region of extreme interstellar extinction. The central region of the Milky Way has undergone multiple episodes of star formation, resulting in old, intermediate age, and young populations (e.g., Sch\"odel et al. \citeyear{schodel2020}; Nogueras-Lara et al. \citeyear{nogueras-lara2021}) having a range of metallicities (Rich et al. \citeyear{rich2017}; Thorsbro et al. \citeyear{thorsbro2020}; Feldmeier-Krause et al. \citeyear{feldmeier-krause2017}; Bentley et al. \citeyear{bentley2022}).  
Although challenging to observe, due to both crowding and extinction, the NSC offers a valuable stellar population in which to probe both stellar and chemical evolution in a, perhaps, unique environment in the Milky Way.

A recent star formation episode in the NSC, with an age of $\sim$10$^{7}$ yr, resulted in a collection of both hot and cool massive supergiants, with a number of O and B stars, along with WR stars comprised of both WN and WC types (Paumard et al. \citeyear{paumard2006}; Feldmeier-Krause et al. \citeyear{feldmeier-krause2015}).  One member of this young population of stars is the cool M2 supergiant IRS 7, which is the most luminous star in the Galactic center.  IRS 7 was the first star in the Galactic center in which photospheric chemical abundances were derived by Carr et al. (\citeyear{carr2000}), who found IRS 7 to have an approximately solar metallicity, with [Fe/H]=$-0.02\pm$0.13.  An additional study of Galactic center stars by Ramirez et al. \citeyearpar{ramirez2000} included IRS 7 and derived [Fe/H]$=+0.09$, or A(Fe)=log(N$_{Fe}$/N$_{H})+12.0=7.54$.  A later abundance analysis of several luminous cool stars in the Galactic center by Cunha et al. \citeyearpar{cunha2007} also included IRS 7 and they found [Fe/H]=+0.08, where a solar Fe abundance of A(Fe)$=7.45$ was adopted, so A(Fe)$=7.53$.  IRS 7 was one of two cool Galactic center supergiants analyzed in detail by Davies et al. \citeyearpar{davies2009}, who also derived an Fe abundance slightly above solar, with [Fe/H]$=+0.13$. Taken together, all four studies based upon high-resolution spectroscopy yield a consistent Fe abundance of [Fe/H]$=+0.07\pm0.06$ (A(Fe)$=7.52$) for IRS 7.

IRS 7 (2MASS 17454004-2900225) is not only interesting as one of the stars first used to anchor the metallicity of a young star in the Galactic center, but also because of its extreme abundances of carbon, nitrogen, and oxygen relative to Fe, as first found by Carr et al. \citeyearpar{carr2000}.  That study derived [$^{12}$C/Fe]$=-0.68$, [$^{14}$N/Fe]$=+1.06$, and [$^{16}$O/Fe]$=-0.62$, and noted that the large depletions of carbon-12 and oxygen-16, in conjunction with the extreme enhancement of nitrogen-14, was the signature of the deep convective dredge-up of interior matter that had been processed through the full H-burning CNO cycles.  Although the abundance signature of modest depletions of $^{12}$C (typically $\sim$-0.1 to -0.4 dex), moderate enhancements of $^{14}$N (typically $\sim$+0.4 - +0.6 dex), and no measurable changes in $^{16}$O, due to red giant first dredge-up have been observed in many abundance studies (e.g., Masseron \& Gilmore \citeyear{masseron2015}; Martig et al. \citeyear{martig2016}; Hasselquist et al. \citeyear{hasselquist2019}; Takeda et al. \citeyear{takeda2019}), such quantitative studies in massive cool supergiants are rare (e.g., Lambert, Dominy \& Sivertsen \citeyear{lambert1980}; Carr et al. \citeyear{carr2000}).  The mass of IRS 7 has been estimated by the previous spectroscopic studies to be of 17 M$_{\odot}$, 22 M$_{\odot}$, or, 25 M$_{\odot}$ (Carr et al. \citeyear{carr2000}; Ramirez et al. \citeyear{ramirez2000}; Cunha et al. \citeyear{cunha2007}; Davies et al. \citeyear{davies2009}; Tsuboi et al. \citeyear{tsuboi2020}), thus detailed abundance determinations for this massive supergiant are of interest from the perspective of stellar evolution.

IRS 7 became the focus of this study as this star is significantly more massive than the NSC members studied previously by Guer\c{c}o et al. \citeyearpar{guerco2022}, who derived $^{19}$F abundances in the NSC, thus offering the possibility to probe, for the first time, the fluorine abundance in a star displaying the signature of deeply mixed CNO-cycle material. In this study, abundances of mixing sensitive isotopes and elements, in particular fluorine, in IRS 7 are compared to models of massive stars, both standard (i.e., non-rotating) and rotating framework, in order to investigate possible differences between the model prediction and the derived abundances, as well as possibly shed light on the exact evolutionary state of this red supergiant star. 

This paper is organized as follows: in Section \ref{sec:observations} we discuss the observations, in Section \ref{sec:abundances} we present the abundance analysis along with the abundance uncertainties, and the results are discussed in Section \ref{sec:discussion}.

\section{Observations} \label{sec:observations}

High-resolution near-infrared (NIR) spectra of IRS 7 were obtained with the  10-meter Keck II Telescope and the NIRSpec spectrograph at a resolution R = $\lambda$/$\Delta\lambda$ = 25,000. The NIRSpec spectra analyzed have S/N$\sim$120 and span the wavelength range between $\lambda$23,109--23,435\AA, containing the rotational-vibrational (v-r) transitions of H$^{19}$F(1-0) R9 and R11, which were used in the fluorine abundance analysis. This spectral region also contains a number of $^{12}$C$^{16}$O and $^{12}$C$^{17}$O v-r lines that were used to determine the $^{12}$C abundance, as well as the $^{16}$O/$^{17}$O ratio (Section \ref{sec:abundances}).

In addition to the NIRSpec spectrum, a publicly available SDSS DR17 (Abdurro'uf et al. \citeyear{abdurro2022}) APOGEE spectrum of IRS (S/N$\sim$130) was analyzed using the APOGEE line list (Smith et al. \citeyear{smith2021}) to synthesize spectra to determine abundances of the elements C, N, O and Fe. 
IRS 7 was observed as part of the SDSS APOGEE survey (Majewski et al. \citeyear{majewski2017}) and the observations were taken with the APOGEE North spectrograph. The APOGEE spectra have a nominal resolving power R$=\lambda$/$\delta\lambda$=22,500 and spectral coverage between $\sim\lambda$15,150--16,950\AA. 
We note that although there are results for stellar parameters and chemical abundances for IRS 7 in the APOGEE DR17 database, the APOGEE automatic stellar parameters and abundances pipeline ASPCAP (Garc\'ia P\'erez et al. \citeyear{garcia_perez2016}) cannot properly model the IRS 7 spectrum, given the very high macroturbulent velocity of IRS 7, which corresponds to $\sim$25 km/s and is not accounted for in the APOGEE synthetic stellar libraries (Zamora et al. \citeyear{zamora2015}). 

\section{Abundance Analysis and Uncertainties} \label{sec:abundances}

Abundances of C, N, O, and F were derived from molecular transitions and, as part of the spectrum synthesis calculations, molecular equilibria were computed for all significant C-, N-, O-, and F-containing molecules, along with four stages of atomic ionization. 
The spectrum synthesis code, Turbospectrum (Alvarez \& Plez \citeyear{Alvarez_Plez1998}; Plez \citeyear{plez2012}), includes dozens of molecules, with some of the more important ones for this analysis being H$_{2}$, CO, OH, CN, CH, NO, NH, C$_{2}$, N$_{2}$, O$_{2}$, HF, H$_{2}$O, CO$_{2}$, as well as many less important species, such as SiO, SiC, SiN, TiO, HCN, or C$_{3}$.  
The methodology for using specific molecular lines, such as those from CO, OH, or CN, in oxygen-rich red giants (i.e., C/O$<$1) is described in Smith et al. \citeyearpar{smith2013} for the NIR CO, OH, and CN lines, but we recap here; in cool red giants, such as IRS 7, CO locks up most of the carbon, due to its high dissociation energy, making CO the best monitor of the total carbon abundance.  
The CO line strengths are mildly sensitive to the oxygen abundance, while the OH lines are the best indicator of the oxygen abundance. 
Typically one iteration of deriving the C-abundance from CO, with an initial value of the O-abundance, and then using the OH lines, with the initially derived CO carbon abundance to determine an O-abundance and then re-derive the carbon abundance, results in consistent abundances of both C and O.  
With defined abundances of C and O, the nitrogen abundance is then determined from the CN lines.

In Table \ref{tab:linelist} we provide the list of molecular and atomic lines used in the abundance analyses of the elements; $^{12}$C is derived from $^{12}$C$^{16}$O lines, $^{14}$N from $^{12}$C$^{14}$N lines, $^{16}$O from $^{16}$OH lines, $^{19}$F from H$^{19}$F lines, as well as the minor isotopes $^{13}$C from $^{13}$C$^{16}$O lines and $^{17}$O from $^{12}$C$^{17}$O lines. Atomic lines of Fe I were used to derive the metallicity of IRS 7.
The excitation potentials and log gf-values of the transitions are also listed in Table \ref{tab:linelist}, along with the sources of the gf-values given in the last column for each set of lines.

\begin{table*}
\caption{Spectral Lines used in the Abundance Analysis}
\begin{tabular}{lrlcccccc}
\hline \hline
\multicolumn{1}{c}{Atom/Molecule} & \multicolumn{1}{c}{ } & \multicolumn{1}{c}{$\lambda _{air}$} & \multicolumn{1}{c}{$\chi$ } & \multicolumn{1}{c}{$\log$ $gf$} & \multicolumn{1}{c}{Spectrograph} & X & A(X) & gf Source \\
\multicolumn{1}{c}{             } & \multicolumn{1}{c}{ } & \multicolumn{1}{c}{(\AA)           } & \multicolumn{1}{c}{(eV)   } & \multicolumn{1}{c}{           } & \multicolumn{1}{c}{            } &   &      \\
\hline
$^{12}$C$^{14}$N  & \multirow{2}{*}{$\bigg\{$} & 15321.421 & $1.051$\tnote{a}  & $-1.959$\tnote{a} & APOGEE & \multirow{2}{*}{$^{14}$N} & \multirow{2}{*}{8.95} & Sneden et al. \citeyearpar{sneden2014} \\
                  & & 15321.424 & $0.791$\tnote{a}  & $-1.755$\tnote{a} & &          & \\
                  & \multirow{2}{*}{$\bigg\{$} & 15323.433 & $0.864$\tnote{a}  & $-1.876$\tnote{a} & & \multirow{2}{*}{$^{14}$N} & \multirow{2}{*}{8.95} \\
                  & & 15323.806 & $1.204$\tnote{a}  & $-1.534$\tnote{a} & &          & \\
                  & & 15410.558 & $0.840$\tnote{a}  & $-1.555$\tnote{a} & & $^{14}$N & 8.97 \\
                  & & 15447.095 & $1.092$\tnote{a}  & $-1.160$\tnote{a} & & $^{14}$N & 9.05 \\
                  & & 15466.235 & $1.092$\tnote{a}  & $-1.154$\tnote{a} & & $^{14}$N & 8.95 \\
                  & & 15471.812 & $0.860$\tnote{a}  & $-1.708$\tnote{a} & & $^{14}$N & 8.90 \\
                  & & 15481.868 & $0.863$\tnote{a}  & $-1.542$\tnote{a} & & $^{14}$N & 8.95 \\ \hline
$^{12}$C$^{16}$O  & & 23109.370 & $1.510$\tnote{b}  & $-4.907$\tnote{b} & NIRSpec & $^{12}$C & 7.47 & Goorvitch et al. \citeyearpar{goorvitch1994} \\
                  & & 23122.097 & $1.550$\tnote{b}  & $-4.900$\tnote{b} &  & $^{12}$C & 7.49 \\
                  & \multirow{2}{*}{$\bigg\{$} & 23303.581 & $1.421$\tnote{b}  & $-4.517$\tnote{b} & & \multirow{2}{*}{$^{12}$C} & \multirow{2}{*}{7.73} \\
                  & & 23304.591 & $0.485$\tnote{b}  & $-4.994$\tnote{b} &  &  & \\
                  & \multirow{2}{*}{$\bigg\{$} & 23373.021 & $0.396$\tnote{b}  & $-5.124$\tnote{b} &  & \multirow{2}{*}{$^{12}$C} & \multirow{2}{*}{7.73} \\
                  & & 23374.033 & $1.657$\tnote{b}  & $-4.455$\tnote{b} &  & & \\ \hline
$^{13}$C$^{16}$O  & & 16121.212 & $0.543$           & $-6.959$          & APOGEE & $^{12}$C/$^{13}$C & 6.0 & Li et al. \citeyearpar{li2015} \\
                  & & 16121.279 & $0.528$           & $-6.977$          &  & $^{12}$C/$^{13}$C & 6.2 \\
                  & & 16121.413 & $0.559$           & $-6.942$          &  & $^{12}$C/$^{13}$C & 6.2 \\
                  & & 16121.614 & $0.512$           & $-6.995$          &  & $^{12}$C/$^{13}$C & 5.7 \\
                  & & 16121.882 & $0.576$           & $-6.924$          &  & $^{12}$C/$^{13}$C & 5.3 \\
                  & & 16122.215 & $0.498$           & $-7.014$          &  & $^{12}$C/$^{13}$C & 4.2 \\
                  & & 16122.619 & $0.593$           & $-6.907$          &  & $^{12}$C/$^{13}$C & 4.0 \\
                  & & 16123.083 & $0.483$           & $-7.033$          &  & $^{12}$C/$^{13}$C & 4.8 \\
                  & & 16123.625 & $0.611$           & $-6.891$          &  & $^{12}$C/$^{13}$C & 5.5 \\
\hline
$^{12}$C$^{17}$O  & & 23301.588 & $0.230$           & $-5.475$          & NIRSpec & $^{16}$O/$^{17}$O & 800 & Goorvitch et al. \citeyearpar{goorvitch1994} \\
                  & & 23318.533 & $0.202$           & $-5.509$          &         & $^{16}$O/$^{17}$O & 400 \\
                  & & 23327.600 & $0.188$           & $-5.526$          &         & $^{16}$O/$^{17}$O & 300 \\
                  & & 23337.062 & $0.175$           & $-5.545$          &         & $^{16}$O/$^{17}$O & 600 \\
\hline
Fe I              & & 15207.526 & $5.385$           & $+0.067$           &  APOGEE       & Fe & 7.53 & Smith et al. \citeyearpar{smith2021} \\
                  & & 15395.718 & $5.621$           & $-0.393$           &         & Fe & 7.51 \\
                  & & 15490.337 & $2.198$           & $-4.759$           &         & Fe & 7.35 \\
                  & & 15531.752 & $5.642$           & $-0.564$           &         & Fe & 7.45 \\
                  & & 15648.510 & $5.426$           & $-0.668$           &         & Fe & 7.52 \\
                  & & 15964.865 & $5.921$           & $-0.140$           &         & Fe & 7.43 \\
                  & & 16040.654 & $5.874$           & $+0.071$           &         & Fe & 7.43 \\
                  & & 21238.467 & $4.956$\tnote{b}  & $-1.300$\tnote{b} & NIRSpec & Fe & 7.75 & Guer\c{c}o et al. \citeyearpar{guerco2019} \\ 
                  & \multirow{2}{*}{$\bigg\{$} & 21284.344 & $5.669$\tnote{b}  & $-3.320$\tnote{b} &  & \multirow{2}{*}{Fe} & \multirow{2}{*}{7.75} \\
                  & & 21284.348 & $3.071$\tnote{b}  & $-4.470$\tnote{b} & &    & \\
                  & & 22473.268 & $6.119$\tnote{b}  & $+0.381$\tnote{b} &  & Fe & 8.00 \\ \hline
H$^{19}$F         & & 23134.757 & $0.332$\tnote{c}  & $-3.942$\tnote{c} & NIRSpec & $^{19}$F  & 3.65 & J\"onsson et al. \citeyearpar{jonsson2014a} \\
                  & & 23358.329 & $0.227$\tnote{cd} & $-3.962$\tnote{c} & & $^{19}$F  & 3.73 \\ \hline
$^{16}$OH         & \multirow{2}{*}{$\bigg\{$} & 15391.057 & $0.494$\tnote{a}  & $-5.512$\tnote{a} & APOGEE & \multirow{2}{*}{$^{16}$O} & \multirow{2}{*}{8.30} & Brooke et al. \citeyearpar{brooke2016} \\
                  & & 15391.205 & $0.494$\tnote{a}  & $-5.512$\tnote{a} &  &  &  \\
                  & & 15568.782 & $0.299$\tnote{a}  & $-5.337$\tnote{a} & & $^{16}$O & 8.35 \\
                  & \multirow{3}{*}{$\Bigg\{$} & 15627.289 & $0.886$\tnote{a}  & $-5.514$\tnote{a} & & \multirow{3}{*}{$^{16}$O} & \multirow{3}{*}{8.30} \\
                  & & 15627.291 & $0.886$\tnote{a}  & $-5.514$\tnote{a} & & & \\
                  & & 15627.405 & $0.541$\tnote{a}  & $-5.268$\tnote{a} & & & \\ \hline
\label{tab:linelist}
\end{tabular}
\end{table*}

Synthetic spectra for IRS 7 were computed using the Turbospectrum code (Alvarez \& Plez \citeyear{Alvarez_Plez1998}; Plez \citeyear{plez2012}) and the BACCHUS wrapper (Masseron, Merle \& Hawkins \citeyear{masseron2016}); a Gaussian broadening was adopted to match the  macroturbulent velocity of $\sim25$ km/s. The stellar parameters for IRS 7 were taken from the previous study by Cunha et al. \citeyearpar{cunha2007}: T$_{\rm eff}=3650$ K, log g$=-0.50$, microturbulent velocity $\xi=3.2$ km/s (see below), noting that these are in general agreement with earlier results by Carr et al. (\citeyear{carr2000}; T$_{\rm eff}=3470 $K; log g$=-0.60$; $\xi=3.3$ km/s) and by Davies et al. (\citeyear{davies2009}; T$_{\rm eff}=3600$ K; log g$=0.00$; $\xi=3.0$ km/s). 
A `tailored' MARCS spherical model atmospheres (Gustafsson et al. \citeyear{gustafsson2008}) was computed for the adopted stellar parameters of IRS 7, for a mass of M$=25$ M$_{\odot}$ (Davies et al. \citeyear{davies2009}), a metallicity [Fe/H]$=+0.1$, and a chemical composition in general agreement with the abundances for C, N, O derived in this study (Table \ref{tab:abundances}), while abundances of other elements were set to solar-scaled.

Molecular dissociation energies and partition functions in Turbospectrum are identical to those adopted in the MARCS model atmosphere code (Gustafsson et al. \citeyear{gustafsson2008}).  New data are evaluated, with updates being made over the years.  In the case of CN, the current dissociation energy used is  7.724$\pm$0.017eV from Ruscic \& Bross \citeyearpar{ruscic2021}; the method used is described in Ruscic et al. \citeyearpar{ruscic2004} and Ruscic et al. \citeyearpar{ruscic2005}, and is derived via a global optimised solution to a large thermochemical network in an error-weighted space of all determinations of thermochemical values (essentially the enthalpies of formation).

\begin{table*}
\centering
\caption{IRS 7 Abundances \& Abundance Sensitivities to Stellar Parameters}
\begin{tabular}{ccccccc}
\hline \hline
X & $<$A(X)$>$  & $\delta$T$\rm_{eff}$ & $\delta\log g$ & $\delta$[Fe/H] & $\delta\xi$   & $\Delta$A$^{a}$  \\ 
  & (This work) &  +100K  & +0.25  & +0.1 & +1.0 km$\cdot$s$^{-1}$ & \\ \hline
$^{19}$F  & 3.69 $\pm$ 0.04 & $+0.20$ & $+0.02$ & $+0.05$ & $-0.02$ &  $\pm$ 0.21 \\
$^{12}$C  & 7.61 $\pm$ 0.12 & $+0.05$ & $+0.10$ & $+0.10$ & $-0.11$ & $\pm$ 0.19 \\
$^{14}$N  & 8.96 $\pm$ 0.04 & $-0.11$ & + 0.13 & + 0.05 & $-0.02$ & $\pm$ 0.18 \\
$^{16}$O  & 8.32 $\pm$ 0.02 & +0.20 & $-0.03$ & + 0.11 & $-0.05$ & $\pm$ 0.23 \\
Fe        & 7.57 $\pm$ 0.19 & $-0.01$ & $+0.04$ & $+0.01$ & $-0.22$ & $\pm$ 0.22 \\
$^{12}$C/$^{13}$C        & 5.3 $\pm$0.8   & ...  & ...       & ....& $-1.2$ & $\pm$ 1.2   \\
$^{16}$O/$^{17}$O        & 525 $\pm$ 220  & ... & ...       & ... & $-120$ & $\pm$ 120   \\
\hline
\end{tabular}
\begin{tablenotes}
\item \textbf{Notes}: Baseline model: T$\rm _{eff}$ = 3650 K; log g = --0.50; [Fe/H] = 0.12; $\xi$ = 3.20 km/s; M = 25 M$_\odot$. (a) Defined as $\Delta$A $(\delta T_{\rm eff})^2 + (\delta \log g)^2 + (\delta [Fe/H])^2 + (\delta \xi)^2]^{1/2}$.
\end{tablenotes}
\label{tab:abundances}
\end{table*}

The microturbulent velocity, $\xi$, for IRS 7 was estimated initially following the method used by Guer\c{c}o et al. \citeyearpar{guerco2022} for other NSC stars, which relied on  $^{12}$C$^{16}$O lines at $\lambda2.3 \mu$m spanning a range of line strengths (equivalent widths).
The best value of $\xi$ was that which yielded the same $^{12}$C abundances from both the stronger and weaker CO lines and, for IRS 7, a microturbulent velocity $\xi=4.5$ km-s$^{-1}$ was obtained.
As noted and discussed by Guer\c{c}o et al. \citeyearpar{guerco2022}, the strong molecular lines from CO or OH tend to require larger microturbulent velocities when compared to typically weaker and higher excitation atomic or molecular lines. 
The synthetic fits to the APOGEE spectra near $\lambda1.6\mu$m, however, were overall better when adopting lower values of microturbulent velocities of $\xi\sim2 $ km-s$^{-1}$. Given the range in microturbulent velocity values required across a range of line strengths and wavelength regions, the adopted microturbulent velocity in all calculations was $\xi$=3.2$\pm$1.0 km-s$^{-1}$, which we note, is the microturbulent velocity parameter derived in Cunha et al. \citeyearpar{cunha2007} and also similar to those values in previous studies of IRS 7 (Carr et al. \citeyear{carr2000}; Davies et al. \citeyear{davies2009}). 

\begin{figure}
\centering
\includegraphics[width=\columnwidth]{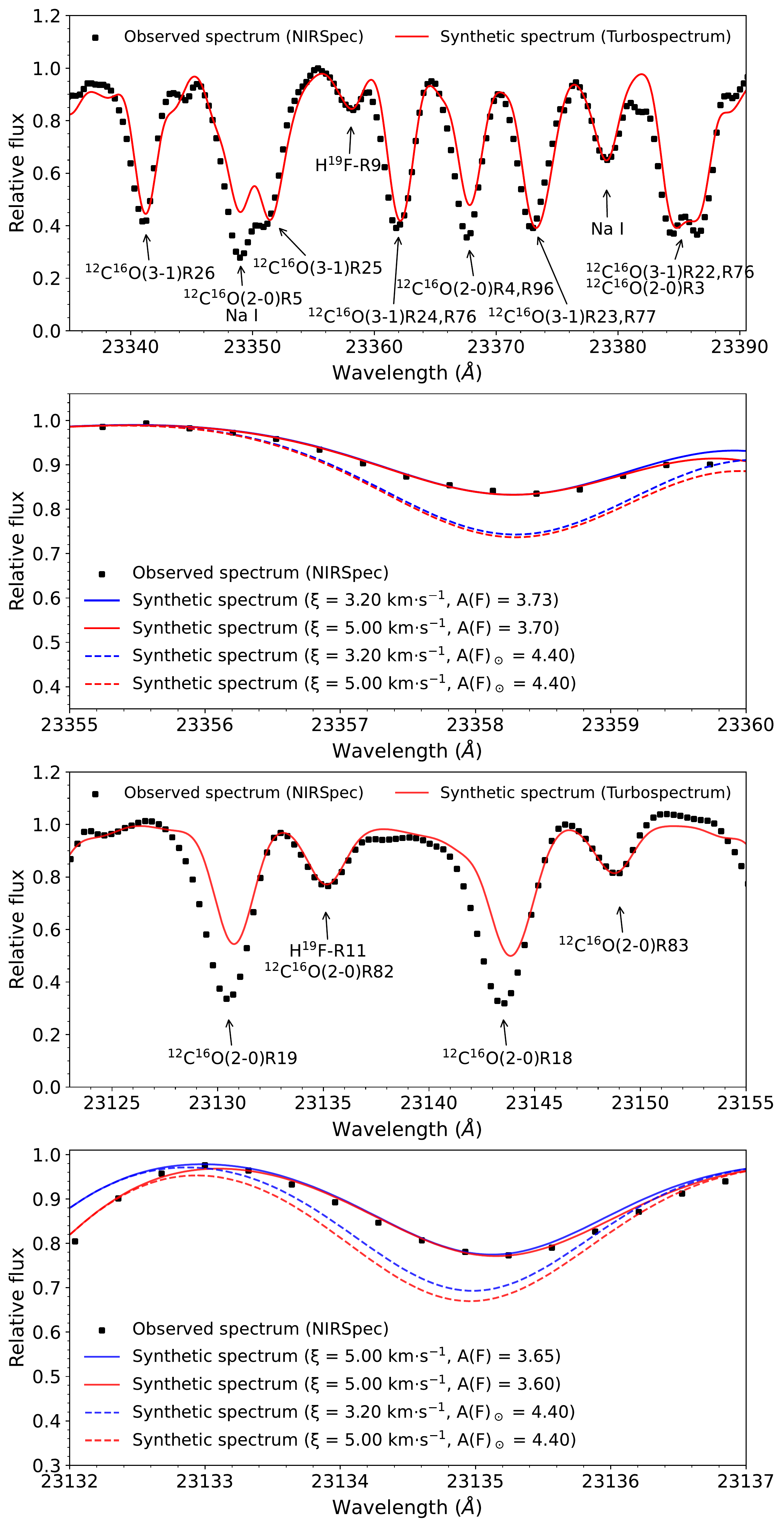}
\caption{Observed (black squares) and synthetic spectra (solid and dashed curves) of the star IRS 7 showing an extended region around the HF (1--0) R9 line (top panel); a zoom of the HF R9 line (second panel from top), and the region around the HF R11 line (third panel from top) and a zoom of the R11 line. The red syntheses were computed for $\xi$ = 5.00. We also show two best fitting syntheses computed for $\xi$ = 3.20 km$\cdot$s$^{-1}$ and A(F) = 3.73 (blue solid line) and for $\xi$ = 5.00 km$\cdot$s$^{-1}$ and A(F)=3.70 (red solid line). The small difference between these two syntheses illustrates the insignificant effect that the microtubulent velocity parameter has on the derived fluorine abundance. We also show additional syntheses for solar fluorine (A(F)$_\odot$ = 4.40; Maiorca et al. \citeyear{maiorca2014}) using $\xi$ = 3.20 km$\cdot$s$^{-1}$ (blue dashed line) and $\xi$ = 5.00 km$\cdot$s$^{-1}$ (red dashed line); this illustrates clearly that solar fluorine abundance does not fit the observed spectrum of IRS 7.}
\label{fig:synthesis}
\end{figure}

The main focus of this study is fluorine and the two HF lines analyzed in IRS 7 are weak and derived $^{19}$F abundances are quite insensitive to the microturbulent velocity; the individual HF line abundances are in Table \ref{tab:linelist}.
In Figure \ref{fig:synthesis} are shown observed and synthetic spectra for the HF(1-0) R9 and R11 lines, with filled black squares being the observed IRS 7 spectrum and the continuous red and blue curves being synthetic spectra as labelled in the figure key.  The top two panels highlight the R9 line, while the bottom two panels show the R11 line.  The top (first) panel plots a 55\AA-wide region around the R9 line to illustrate the overall absorption-line spectrum of IRS 7, while the second panel provides a close-up 5\AA\ view of the R9 line profile.  Compared to the R11 line (discussed below), that is heavily blended with a $^{12}$C$^{16}$O line, the R9 line is relatively clean, being blended only with the $^{12}$C$^{17}$O(2-0) R25 line, which falls $\sim$1.2\AA\ to the blue and is quite weak in IRS 7.  The analysis of the $^{12}$C$^{17}$O lines is discussed below, but the $^{17}$O abundance that is derived from the $^{12}$C$^{17}$O(2-0) R27, R28, R29, and R31 lines (with $^{16}$O/$^{17}$O=525) is used in the syntheses shown in the top two panels of Figure \ref{fig:synthesis}. A striking aspect of the R9 line in IRS 7 is its weakness relative to other red giant members of the NSC having similar effective temperatures (e.g., the top panel of Figure \ref{fig:synthesis} in Guer\c{c}o et al. \citeyearpar{guerco2022}).  The weakness of the R9 line renders it insensitive to the microturbulent velocity and this point is illustrated in the second panel of Figure \ref{fig:synthesis}, where synthetic spectra having $\xi$=3.2 km-s$^{-1}$ (blue) and 5.0 km-s$^{-1}$ (red) are shown for two $^{19}$F abundances.  The lower fluorine abundance of A(F)=3.73 is favored over A(F)=4.40 (solar; Maiorca et al. \citeyear{maiorca2014}), independent of $\xi$, indicating that IRS 7 has a significantly lower abundance of $^{19}$F than other NSC red giants (where A(F)$\sim$4.8; Guer\c{c}o et al. \citeyear{guerco2022}).

The spectrum around the R11 line is shown in the third panel of Figure \ref{fig:synthesis} ($\sim$35\AA) and again illustrates the weakness of the HF lines in IRS 7.  As noted above, the R11 line is blended significantly with the $^{12}$C$^{16}$O(2-0) R82 line that falls 0.52\AA\ to the red.  The $^{12}$C$^{16}$O near $\lambda$23149\AA\ is the R83 line, which will be about the same strength as the R82 line; the fact that the blended HF R11 and CO R82 feature is only marginally stronger than the R83 line suggests that HF is not the major contributor to the blend.  This suspicion is born out by the fourth panel in Figure \ref{fig:synthesis}, where the line profile of the HF/CO blend is shown in detail.  As with the R9 line, the lower abundance of A(F)=3.62 provides the best fit and points to a depleted $^{19}$F abundance in IRS 7 relative to other NSC red giants.

One final note about Figure \ref{fig:synthesis} concerns the third panel and the two strong absorption lines that are not fit by the synthesis: these are the low-excitation $^{12}$C$^{16}$O(2-0) R18 ($\chi$=0.08 eV) and R19 ($\chi$=0.09 eV) lines.  The inability of spectrum synthesis using static 1-d model atmospheres to model these very strong, low-excitation molecular lines has been known for some time and is discussed in detail in Tsuji (\citeyear{tsuji1988}; \citeyear{tsuji1991}) and Smith \& Lambert (1986; 1990).  These studies found this effect for the first-overtone v-r CO lines studied here, as well as the fundamental v-r CO lines near $\lambda$4.6$\mu$m, and the fundamental v-r OH lines near $\lambda$3.9$\mu$m, and suggested that the truncation of model atmospheres at low optical depths plays a role, as well as a single value (depth-independent) for the microturbulent velocity.  These studies did find, however, that the use of weaker and typically higher-excitation molecular lines leads to secure abundances.

\begin{figure}
\centering
\includegraphics[width=\columnwidth]{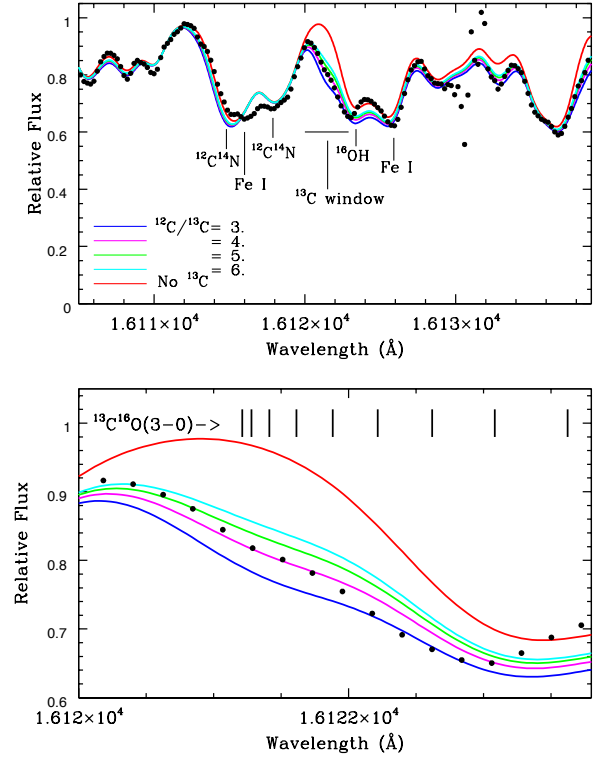}
\caption{The top and bottom panels illustrate synthetic (continuous colored lines) and IRS 7 (black filled circles) spectra centered on the $^{13}$C$^{16}$O(3-0) bandhead, with five synthetic spectra having different carbon-13 abundances (one with no $^{13}$C and four with $^{12}$C/$^{13}$C= 3, 4, 5, and 6). The top panel provides a 35\AA-wide window of the spectral region around the $^{13}$C$^{16}$O bandhead in order to provide a view of the overall absorption in IRS 7 and how this compares to the absorption from $^{13}$C$^{16}$O lines.  The glitch in the IRS 7 spectrum at $\lambda$16130\AA\ is caused by poor telluric subtraction.  The bottom panel is a close-up of the bandhead and illustrates the IRS 7 pixels that are most sensitive to the $^{12}$C/$^{13}$C ratio; the isotopic ratio was set by this spectral interval from $\sim\lambda$6120-6123\AA, resulting in $<^{12}$C/$^{13}$C$>$=5.3$\pm$0.8.  The line positions of the individual $^{13}$C$^{16}$O lines are illustrated by the vertical lines at the top of the bottom panel.  The red end of the $^{13}$C$^{16}$O bandhead is blended with a strong OH line plus an Fe I line (with an uncertain gf-value) further to the red.}
\label{fig:c13}
\end{figure}

The abundances of the minor isotopes $^{13}$C and $^{17}$O, presented as ratios of $^{12}$C/$^{13}$C and $^{16}$O/$^{17}$O, were set by weak lines of $^{13}$C$^{16}$O and $^{12}$C$^{17}$O, respectively.  The NIRSpec wavelength range for IRS 7 did not cover any detectable lines containing carbon-13, so the ratio of $^{12}$C/$^{13}$C was determined from the APOGEE spectrum using the $^{13}$C$^{16}$O(3-0) bandhead near $\sim\lambda$16,120-16,125\AA, which is one of the strongest $^{13}$C features in the APOGEE spectral window in IRS 7.  A $\sim$3\AA\ window covering the $^{13}$C$^{16}$O(3-0) bandhead was used, with the best-fit being that which minimized the residuals between observed and synthetic spectra.  
Comparisons of syntheses with the observed relative fluxes are illustrated in the two panels of Figure \ref{fig:c13}, with five synthetic spectra having different carbon-13 abundances shown in both the top and bottom panels, along with the IRS spectrum (filled black circles): the synthetic spectra are plotted as solid curves of different colors for each value of $^{12}$C/$^{13}$C, or in the case of the red synthesis, a $^{13}$C abundance of zero.  Each synthetic spectrum is calculated with the $^{12}$C abundance derived for IRS 7 (A($^{12}$C)=7.61)  The top panel shows a wider region of the spectrum ($\sim$35\AA) around the $^{13}$C$^{16}$O(3-0) bandhead, in order to illustrate the overall absorption spectrum in this region compared to the absorption from $^{13}$C$^{16}$O itself.  Most of the absorption is from molecules consisting of a mix of CN, CO, and OH lines, along with a smaller number of atomic lines; the major absorption features straddling the $^{13}$C$^{16}$O bandhead window are labelled.  The bottom panel is a zoom covering only the region around the $^{13}$C$^{16}$O bandhead and illustrates the spectral pixels that are most sensitive to the $^{13}$C abundance.  The IRS 7 pixels covering this region were used to set the carbon isotopic ratio, with the mean value and standard deviation being $^{12}$C/$^{13}$C=5.3$\pm$0.8.

Oxygen-17 was detected in the NIRSpec spectra via $^{12}$C$^{17}$O(2-0) lines near $\lambda$ $\sim$23,300\AA, including the R25, R26, R27, R28, R29, R30, and R31 lines, which were used in Guer\c{c}o et al. \citeyearpar{guerco2022} to derive $^{16}$O/$^{17}$O ratios in NSC stars; a number of these $^{12}$C$^{17}$O lines are shown in Figure 3, where the region from $\lambda$23295-23340\AA\ is shown, including the observed IRS 7 spectrum (black filled circles), along with four synthetic spectra having values of $^{16}$O/$^{17}$O=100, 200, 400, and one with no $^{17}$O.  This region illustrates nicely both the weakness of the oxygen-17 isotopic lines, relative to the much stronger $^{12}$C$^{16}$O lines, as well as the blending of these weak lines, with the blending created to a large degree by the large macrotubulent velocity observed in IRS 7. Nevertheless, the oxygen-17 isotope is detected in several lines and the mean abundance from four lines (the R30 line is hopelessly blended with a Fe I line whose gf-value is somewhat uncertain), presented as the isotopic ratio is $^{16}$O/$^{17}$O=525$\pm$220. 

\begin{figure}
    \centering
    \includegraphics[width=\columnwidth]{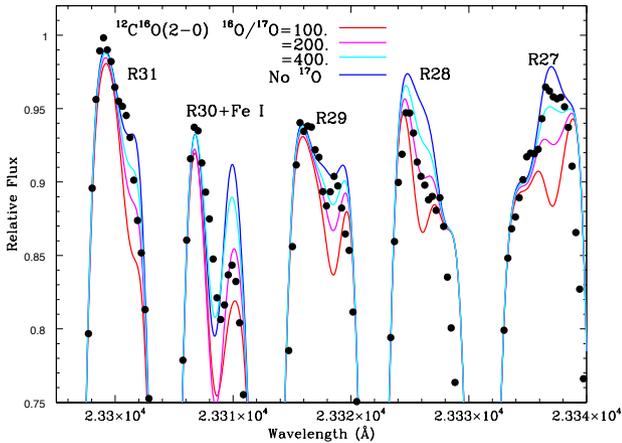}
    \caption{A 45\AA-wide spectral window illustrating the $^{12}$C$^{17}$O lines used to estimate the $^{16}$O/$^{17}$O ratio for IRS 7. Synthetic spectra, with different values of $^{16}$O/$^{17}$O as noted in the plot key, are shown as solid colored curves, with the observed IRS 7 spectrum plotted as solid black circles. This region contains the $^{12}$C$^{17}$O(2-0) R27, R28, R29, R30, and R31 lines, with all of them blended to some degree with a number of strong $^{12}$C$^{16}$O(2-0) and (3-1) lines running through this region.  Due to this significant blending, the derived $^{17}$O abundance (presented as the $^{16}$O/$^{17}$O ratio) in IRS 7 is uncertain, although the $^{17}$O isotopic lines are detected clearly.}
\end{figure}

Table \ref{tab:abundances} presents the mean abundances obtained for F, C, N, O, and Fe plus the corresponding standard deviations computed from the individual line abundance measurements. For most of the studied elements the internal line-to-line abundance variations are less than $\sigma$ $\sim$0.1 dex, except for Fe ($\sigma$ $\sim$0.2 dex). Abundance sensitivities to changes in the parameters T$_{\rm eff}$, log g, metallicity, and microturbulent velocity are also presented in Table 2 along with $\Delta$A, defined as: 
[$(\delta T_{\rm eff})^2 + (\delta \log g)^2 + (\delta [Fe/H])^2 + (\delta \xi)^2]^{1/2}$.
The typical combined error for the adopted parameter variations is $\sim$ 0.25 dex or less. 

The isotopic ratios are set by fits to various isotopic combinations of CO lines, thus their uncertainties are relatively insensitive to stellar atmospheric parameters, but are due mainly to the intrinsic weakness of the $^{13}$C$^{16}$O and $^{12}$C$^{17}$O lines themselves (e.g., Abia et al. \citeyear{abia2012}).  Uncertainties in Table 2 for the minor isotopes were derived based on the scatter in the isotopic ratios derived around spectral intervals dominated by each individual isotopic line (see Table \ref{tab:linelist}); these uncertainties are noted in Table \ref{tab:abundances}.  In addition, tests of the sensitivities of the $^{12}$C/$^{13}$C and $^{16}$O/$^{17}$O ratios to microturbulence were investigated and it was found that both sets of $^{13}$C$^{16}$O and $^{12}$C$^{17}$O lines were insensitive to changes in $\xi$ of $\pm$1.0 km-s$^{-1}$, meaning that these lines are weak (unsaturated).  The sensitivities to the isotope ratios due to changes in $\xi$ are set by the $^{12}$C$^{16}$O lines, or to the derived $^{12}$C abundances as a function of $\xi$: changes to the isotopic ratios with $\Delta\xi$=+1.0 knm-s$^{-1}$ are presented in Table \ref{tab:abundances}.

\section{Discussion} \label{sec:discussion}

\begin{figure*}
\centering
\includegraphics[width=0.75\textwidth]{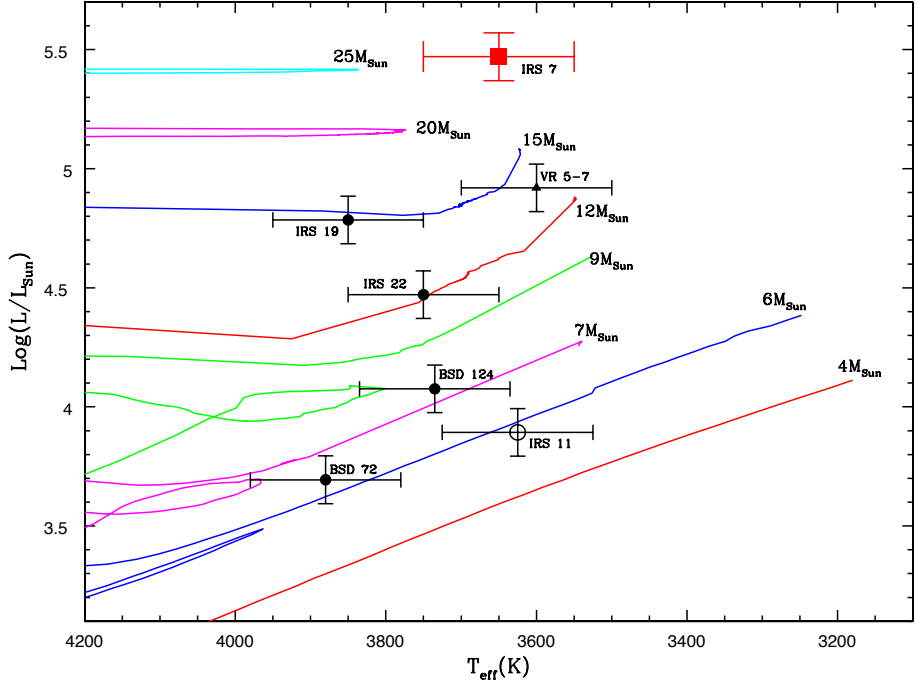}
\caption{Luminosity (as Log(L/L$_{\odot}$)) versus T$_{\rm eff}$ for the Galactic center stars discussed here.  The filled and open circles are all members of the NSC, as is IRS 7, while VR 5-7 (filled triangle) is a member of the Quintuplet Cluster.  The 7M$_{\odot}$, 9M$_{\odot}$, 12M$_{\odot}$, 15M$_{\odot}$, 20M$_{\odot}$, and 25M$_{\odot}$ stellar models are rotating models from Ekstr\"om et al. \citeyearpar{Ekstrom2012}, while the 4 M$_{\odot}$ and 6M$_{\odot}$ stellar models are from Lagarde et al. \citeyearpar{lagarde2012}. This figure presents both the observed stars and stellar models in an evolutionary context within a type of HR Diagram.}
\label{fig:hr}
\end{figure*}

As previously mentioned, Guer\c{c}o et al. \citeyearpar{guerco2022} analyzed a sample of massive young cool and evolved stellar members of the Nuclear Star Cluster, of which IRS 7 is a member. 
Figure \ref{fig:hr} shows an HR diagram (log L versus T$_{\rm eff}$), highlighting the position of the target star IRS 7 (red square), relative to the other NSC targets studied previously (shown as solid circles, with IRS 11 shown as an open circle). Figure \ref{fig:hr} also includes the star VR 5-7 (shown as a triangle), which is a member of the Quintuplet cluster, at a distance of approximately 30 pc from the Galactic center. Models from Ekstr\"om et al. \citeyearpar{Ekstrom2012} are shown for stellar masses 25M$_{\odot}$; 20M$_{\odot}$; 15M$_{\odot}$; 12M$_{\odot}$; 9M$_{\odot}$ and 7M$_{\odot}$; the stellar models shown are solar metallicity rotating models.
Similar models, but for higher metallicity, are found in the recent work by Yusof et al. \citeyearpar{yusof2022}. In Figure \ref{fig:hr} we also show two models corresponding to lower mass stars of 6M$_{\odot}$ and 4M${_\odot}$ from Lagarde et al. \citeyearpar{lagarde2012}.  
In the log L-T$_{\rm eff}$ regime shown, the 4M$_{\odot}$ and 6M$_{\odot}$ models from Lagarde et al. \citeyearpar{lagarde2012} span the thermally-pulsing asymptotic giant branch (TP-AGB) phase of stellar evolution, in which the internal structure of the star consists of an inert degenerate C-O core, surrounded by He- and H-burning shells.  The so-called third dredge-up (TDU) occurs during TP-AGB evolution, resulting in surface enhancements of $^{12}$C and the slow neutron-capture (s-process) elements.  Stars more massive than $\sim$8-10M$_{\odot}$ do not undergo TP-AGB evolution, but continue to evolve through various core-burning phases, such as core He-burning and core C-burning (e.g., see the review by Karakas \& Lattanzio \citeyear{karakas2014}).  The boundary between the most massive TP-AGB stars and the lowest-mass core-burning supergiants is not known precisely, but likely lies in the range of $\sim$8-9M$_{\odot}$ (Karakas \& Lattanzio \citeyear{karakas2014}).

Given this background, an examination of Figure \ref{fig:hr} would suggest that IRS 7, IRS 19, VR 5-7, and IRS 22 are well-above the mass limit for TP-AGB stars, while BSD 72 and BSD 124 lie near the core-burning supergiant/TP-AGB boundary, with IRS 11 sitting in a location that may be compatible with it being in the TP-AGB phase of evolution. Based on their fluorine abundances discussed below, we may surmise that only IRS 11 may show evidence of being in the TP-AGB phase.

\subsection{The Baseline Fluorine Abundance in the NSC} \label{sec:baseline}

\begin{figure}
\centering
\includegraphics[width=\columnwidth]{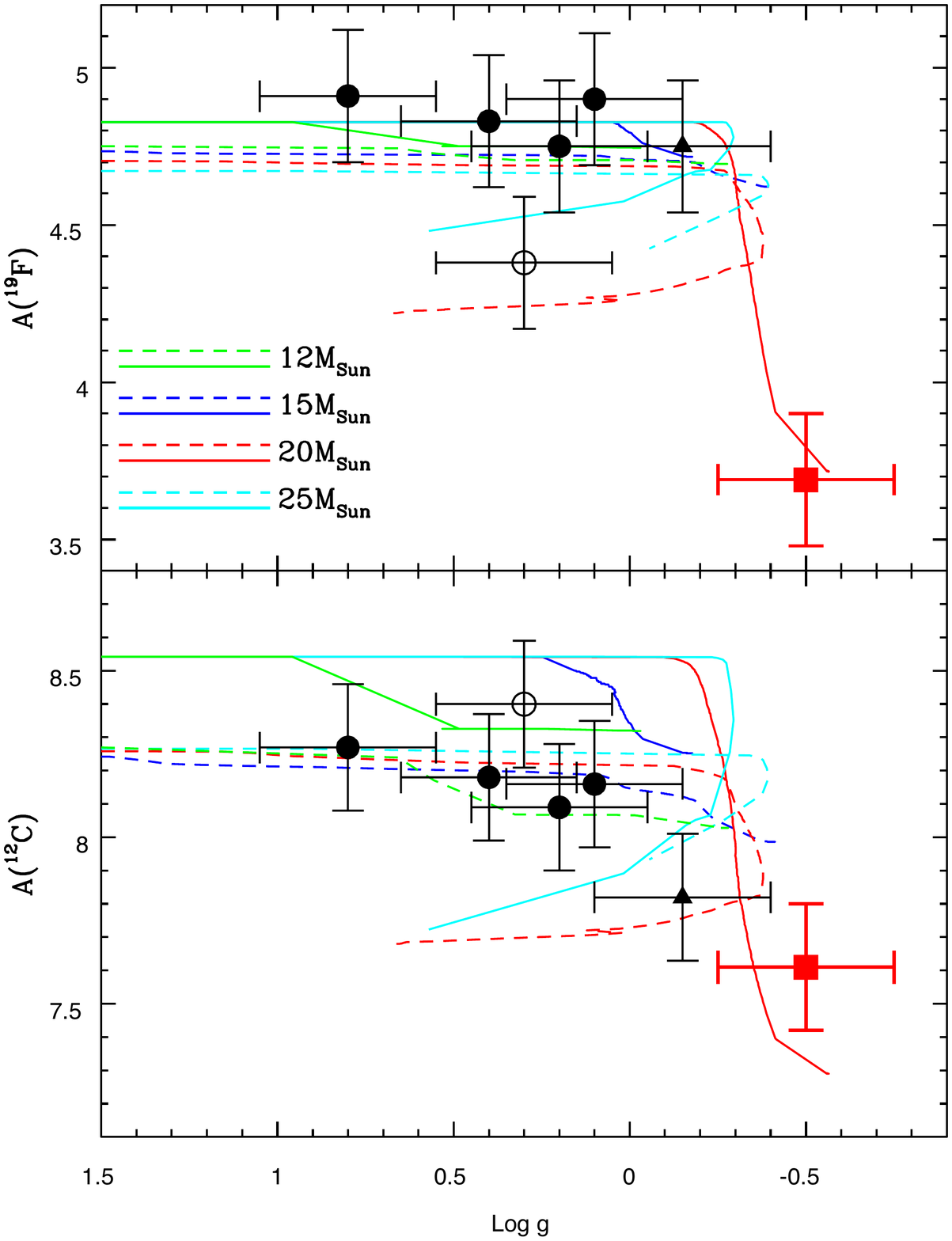}
\caption{The evolution of the fluorine-19 (top panel) and carbon-12 (bottom panel) abundances as a function of the stellar surface gravity. The model curves for $^{12}$C were extracted from data in Ekstr\"om et al. \citeyearpar{Ekstrom2012}, while those for $^{19}$F are presented in this study. The solid lines are non-rotating models and dashed lines are rotating ones. The solid black circles represent stars from the Nuclear Star Cluster from Guer\c{c}o et al. (\citeyear{guerco2022}; open circle is IRS 11); these have estimated masses between $\sim$5--15M$_{\odot}$ and define the baseline abundances for the NSC. IRS 7 (shown as the red square) is more massive (M$\sim$25M$_{\odot}$); its $^{19}$F and $^{12}$C abundances are clearly depleted. VR5-7 from the Quintuplet Cluster (roughly 30 parsecs away from the Galactic center) is also shown (black filled triangle).  The massive rotating models that reverse evolution towards lower surface gravities back to higher values of log g are driven by high mass-loss rates, moving the star towards higher values of T$_{\rm eff}$, smaller radii, and larger surface gravities. 
}
\label{fig:plotlumCF}
\end{figure}

The fluorine abundances of the Galactic center targets from Guerco et al. \citeyearpar{guerco2022} are plotted as a function of surface gravity (log g) in the top panel of Figure \ref{fig:plotlumCF} as black symbols. Surface gravity is plotted along the abscissa, as the relatively massive stars shown in Figure \ref{fig:plotlumCF} evolve off of the main sequence towards both lower effective temperatures and lower surface gravities, with stars evolving towards the right in Figure \ref{fig:hr}. 
The mean fluorine abundance of the NSC stars in the figure is $<$A($^{19}$F)$>$ = 4.75$\pm$0.20, or, $<$[F/Fe]$>=+0.28\pm 0.18$. 

Guer\c{c}o et al. \citeyearpar{guerco2022} pointed out that their sample had fluorine abundances roughly consistent with the A(F) versus A(Fe) trend defined by Galactic thin disk stars having similar metallicities (see their figure 4); the abundance results from disk stars from the literature scattered both above and below the line representing the secondary behavior of A(F) versus A(Fe) in their figure 4. It is noted here, however, that 
all their NSC targets, except for IRS 11 (shown as the open circle in Figure \ref{fig:hr}), had similar fluorine abundances within the uncertainties: an average fluorine abundance and standard deviation of $<$A($^{19}$F)$>$ $=4.85\pm0.06$, being on average $\sim$0.15 dex above the secondary behavior. 

IRS 11 will be discussed further in the next section but concerning its measured fluorine abundance, it should be kept in mind that this was the only star in Guerco et al. \citeyearpar{guerco2022} having only the HF(1-0) R11 line  measured, which is severely blended with a CO line.

It is reasonable to assume that the average fluorine abundance of the studied population of young NSC stars of A(F)=4.75 represents the baseline fluorine abundance of the gas material that formed the young and massive stars in the NSC, and that this baseline abundance has been reached via the chemical evolution in the disk, star formation in the center and the gas falling into the Galactic center.
One particular aspect of the targets in the Guer\c{c}o et al. \citeyearpar{guerco2022} study that is relevant here is that, although massive, those NSC targets previously studied cover a range in mass between $\sim$ 5 - 15M$_{\odot}$, which is significantly lower than the estimated mass of $\sim$25M$_{\odot}$ for IRS 7, the most extreme star in the Galactic center. The fluorine abundance obtained for IRS 7 (A(F)$=3.69$; shown as the red square in the top panel of Figure \ref{fig:hr}) is over 1 dex lower than the baseline NSC fluorine abundance value, being in stark contrast with the other fluorine abundance results for the cluster, and suggesting that its initial fluorine content has been destroyed, as will be discussed in the following section.

\subsection{The Fluorine Abundance, CNO-Cycle Signatures, and Model Predictions}

The low $^{19}$F abundance of IRS 7, relative to the other Galactic center red giants of lower masses, in the top panel of Figure \ref{fig:plotlumCF} is striking and beyond the estimated fluorine abundance uncertainties.  Included in the bottom panel of Figure \ref{fig:plotlumCF} are $^{12}$C abundances for each star as a comparison to $^{19}$F.  Also in both panels of Figure \ref{fig:plotlumCF} are results from the stellar models of Ekstr\"om et al. \citeyearpar{Ekstrom2012}; the predictions for fluorine are presented for the first time here and were computed in the framework of the stellar models from Ekstr\"om et al. \citeyearpar{Ekstrom2012}. 
The models shown cover a mass range of $\sim10-25$ M$_{\odot}$ and include models with 12M$_{\odot}$, 15M$_{\odot}$, 20M$_{\odot}$, and 25M$_{\odot}$, with the solid lines representing non-rotating standard models, while the dashed lines represent rotating stellar models having initial rotations on the ZAMS of between 235 and 306 km s$^{-1}$. These values correspond to time averaged surface velocities during the main-sequence phase of between 178 and 217 km s$^{-1}$ (see table 2 in Ekstr\"om et al. \citeyear{Ekstrom2012}).  The stellar models are plotted with different colors and are labelled.
These models assumed solar abundances (having initial values of X=0.720, Y=0.266, Z=0.014, which account for diffusion), with the heavy-element mixture from Asplund et al. \citeyearpar{asplund2005}, except for neon, which is from Cunha et al. \citeyearpar{cunha2006}.  The initial $^{19}$F abundance for the models in Figure \ref{fig:plotlumCF} were shifted to the mean value of A($^{19}$F)$=4.75$ (Guer\c{c}o et al. \citeyear{guerco2022}) as defined by the NSC stars, as discussed above.
This is an assumed approximation, since the overall Fe abundances are only mildly enhanced, while keeping in mind that the initial stellar hydrogen content was originally higher, as the surface H abundance also becomes mildly depleted in these evolved red supergiants (see discussion in Davies et al. \citeyear{davies2009}), resulting in an iron-to-hydrogen ratio which is higher.

The position of IRS 7 in the A($^{12}$C) versus log g diagram (bottom panel of Figure \ref{fig:plotlumCF}) also shows clear depletion relative to the carbon abundances of the other NSC members.
The lower-mass NSC members (with masses 5 - 15M$_{\odot}$) display mild to moderate depletions in A($^{12}$C) of 0.2 - 0.6 dex, along with no significant measurable changes in $^{19}$F (compared to the depletions of $\sim-0.9$ dex in $^{12}$C and -1.0 dex in $^{19}$F, respectively in IRS 7).  The very large depletions in both the surface $^{12}$C and $^{19}$F abundances exhibited by IRS 7 are predicted by the stellar models to take place at masses of approximately 20M$_{\odot}$ or larger (Figure \ref{fig:plotlumCF}), which fits the various estimates for IRS 7 that cluster around 20 - 25M$_{\odot}$.  The two stars with masses falling just below 20 M$_{\odot}$ are IRS 19 and VR 5-7, with estimated masses of 15M$_{\odot}$ and 14M$_{\odot}$, respectively (Guer\c{c}o et al. \citeyear{guerco2022}).  Both IRS 19 and VR 5-7 have $^{19}$F abundances that fall close to the 15M$_{\odot}$ track (Figure \ref{fig:plotlumCF}, top panel), in accordance with their estimated masses.  IRS 19 also falls near the 15M$_{\odot}$ model track for the $^{12}$C abundance as a function of surface gravity, while VR 5-7 has a $^{12}$C abundance and surface gravity that falls closer to the 20M$_{\odot}$, although the estimated masses are uncertain at levels close to this.

\begin{figure*}
\centering
\includegraphics[width=1.0\textwidth]{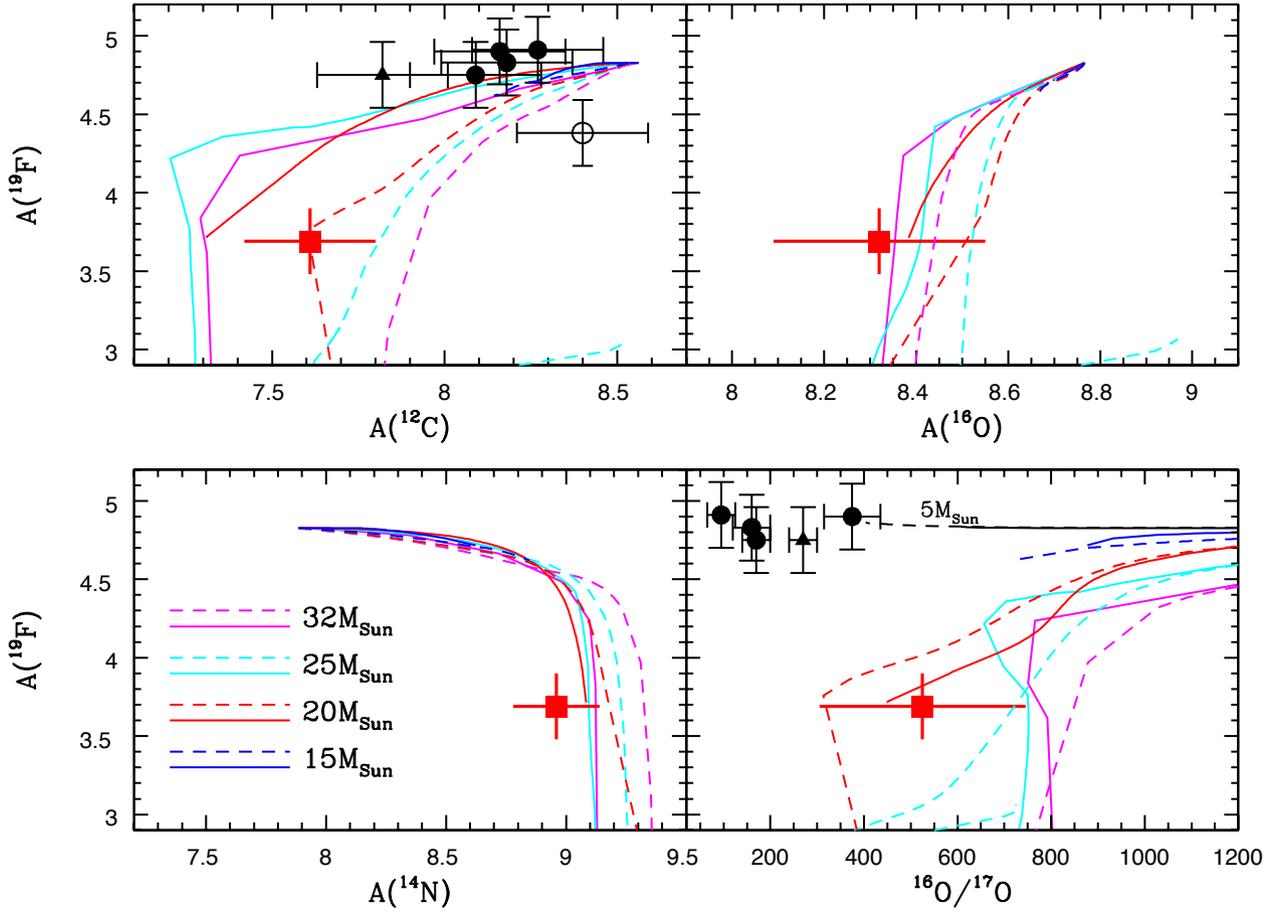}
\caption{The relations of the $^{19}$F abundance with the abundances of $^{12}$C, $^{14}$N, $^{16}$O, and the isotopic ratio $^{16}$O/$^{17}$O. IRS 7 is represented by the red square. The same massive stars in the Galactic center, their corresponding symbols, and models are the same as shown in Figure 3. We also show a 5M$_{\odot}$ model from Ekstr\"om et al. \citeyearpar{Ekstrom2012} as the solid and dashed black curves.}
\label{fig:plotFvsCNO}
\end{figure*}

Additional comparisons between the evolved stars in the NSC and stellar models are shown in Figure \ref{fig:plotFvsCNO}, where various pairs of abundances are plotted in order to further illuminate the mass and evolutionary state of IRS 7.  Pairing mixing-sensitive elements or isotopes together can, in certain cases, increase the comparison power of observationally-derived abundances with predicted abundances from stellar models.  The Galactic center stars discussed here are all young (ages $\sim$10$^{7}$- 10$^{8}$ a) and from either the NSC or the Quintuplet Cluster (VR 5-7), which is only 30 pc from the Galactic center.
Figure \ref{fig:plotFvsCNO} (top left panel) displays the $^{19}$F versus $^{12}$C abundances derived in both IRS 7 and the other Galactic center red giants from Guer\c{c}o et al. \citeyearpar{guerco2022}. 
The observed position of IRS 7 in the $^{12}$C - $^{19}$F plane, relative to the models, is suggestive of the 25 M$_{\odot}$ rotating models providing overall a better prediction of the mixing process(es) that alter the surface abundances of both $^{19}$F and $^{12}$C.
At the position of IRS 7, for a given level of $^{12}$C depletion, the rotating models predict a larger $^{19}$F depletion when compared to the non-rotating models, due to rotational mixing distributing $^{19}$F-poor matter throughout a larger mass volume of the stellar interior.
We note that in the case of the 32M$_{\odot}$ model, there is an increase of both fluorine and carbon at the end (Wolf-Rayet phase), which comes from the fact that layers reflect the nucleosynthesis associated with the beginning of the core He-burning phase. 

Previous approximate estimates from the spectroscopic studies noted in the introduction range from 17 - 25M$_{\odot}$ and, taken at face value, a comparison of the IRS 7 $^{19}$F and $^{12}$C abundances in Figure \ref{fig:plotFvsCNO} falls in-between the rotating 25M$_{\odot}$ and 32M$_{\odot}$ model tracks. These comparisons remain superficial, as the stellar models have not been tailored to the pristine chemical abundances of the NSC, nor the observed rotational velocities of massive main-sequence stars of the NSC. Nonetheless, the top left panel of Figure \ref{fig:plotFvsCNO} presents an insightful window into the interplay between models and the evolutionary state of IRS 7.

The minor isotope of carbon, $^{13}$C, was measured in IRS 7, with $^{12}$C/$^{13}$C$=5.3\pm0.8$, although a plot of A($^{19}$F) versus $^{12}$C/$^{13}$C is not included in Figure \ref{fig:plotFvsCNO}, as $^{12}$C/$^{13}$C does not discriminate between the various stellar models that best match IRS 7. By the time that depleted $^{19}$F has been mixed to the surface of the red supergiant, all models predict a near CN-equilibrium value of $^{12}$C/$^{13}$C$\sim4$, which is within the errors of what is derived here for IRS 7.

Figure \ref{fig:plotFvsCNO} (bottom left panel) adds the $^{14}$N abundance as a comparison to the $^{19}$F abundance in IRS 7.  Since Guer\c{c}o et al. \citeyearpar{guerco2022} did not derive nitrogen abundances in the other Galactic center stars, only IRS 7 is plotted in this panel; again the initial $^{14}$N abundance for IRS 7 is taken to be scaled solar from [Fe/H]$=+0.08$, or an initial abundance of A($^{14}$N)$=7.91$. All of the stellar models predict that no significant $^{19}$F depletion will occur up to $^{14}$N overabundances of $\sim +0.6$ dex; however, by the point when [$^{14}$N/Fe]$\sim+1.0$ dex and greater, the $^{19}$F abundance declines precipitously as $^{14}$N increases. The massive rotating stellar models predict somewhat larger $^{14}$N abundances at a given $^{19}$F depleted abundances by +0.2 - +0.3 dex. The derived $^{14}$N abundance in IRS 7 falls slightly below the values predicted by both the standard and rotating models.  Given that nitrogen abundances are determined from CN, which are also dependent on the derived carbon abundances, the $^{14}$N abundance has a slightly larger uncertainty due to uncertainties in the carbon abundances.

Moving on to the oxygen abundance of IRS 7, Figure \ref{fig:plotFvsCNO} (top right and bottom panels) presents the $^{19}$F abundances as functions of $^{16}$O and the $^{16}$O/$^{17}$O isotopic ratios, respectively.  Again, Guer\c{c}o et al. \citeyearpar{guerco2022} did not measure oxygen abundances in the other Galactic center stars, so only IRS 7 is plotted here, and the initial $^{16}$O abundance is shifted up by 0.08 dex, so the initial oxygen-16 abundances in the NSC is taken to be A($^{16}$O)$=8.77$.
IRS 7 continues to occupy a position in the A($^{19}$F)-A($^{16}$O) plane that points to the signature of full CNO-cycle H-burning, with significant depletion of $^{16}$O accompanying the depleted $^{19}$F abundance. The discriminatory separation between the non-rotating versus rotating models is not as large for $^{16}$O as it is for $^{12}$C, so the derived $^{16}$O abundance and its error, merely verifies the fact that IRS 7 represents a massive red supergiant.

Measurements of the $^{16}$O/$^{17}$O isotopic ratios can help in further probing the conditions for deep mixing in massive stars, in general, and in IRS 7, in particular;
keeping in mind, however, that the destruction of $^{17}$O depends on the cross section of the $^{17}$O(p,$\alpha$)$^{14}$N and $^{17}$O(p,$\gamma$)$^{18}$F reactions, which are uncertain (e.g., see Figure 3 in Lebzelter et al. \citeyear{lebzelter2015}).
As discussed in Section 3, $^{12}$C$^{17}$O(2-0) lines are detectable near the H$^{19}$F(1-0) R9 line, allowing for the derivation of the $^{16}$O/$^{17}$O ratio in several of the Galactic center red giants, including IRS 7.
In this study we obtained $^{16}$O/$^{17}$O$=550$ for IRS 7 (Table \ref{tab:abundances}), while Guer\c{c}o et al. \citeyearpar{guerco2022} derived lower values of $^{16}$O/$^{17}$O for the less massive Galactic center stars they analyzed. 
Comparing with other galactic stars, similar and higher values for this ratio were obtained, for example, for the red supergiants $\alpha$ Ori ($^{16}$O/$^{17}$O = 525) and $\alpha$ Sco ($^{16}$O/$^{17}$O = 850) in Harris \& Lambert \citeyearpar{harris1984}.
The bottom right panel of Figure 4 presents A($^{19}$F) versus $^{16}$O/$^{17}$O for IRS 7 and the other Galactic center stars, along with results from the Ekstr\"om et al. \citeyearpar{Ekstrom2012} stellar models.  The initial $^{16}$O/$^{17}$O ratio for the models is the solar value of 2750, which falls off of the plot scale.  The higher mass stellar models of 20M$_{\odot}$, 25M$_{\odot}$, and 32M$_{\odot}$, all of which deplete $^{19}$F, straddle the observed values of $^{19}$F and $^{16}$O/$^{17}$O derived for IRS 7.  In terms of its fluorine abundance and $^{16}$O/$^{17}$O ratio, IRS 7 falls between the rotating 20M$_{\odot}$ and 25M$_{\odot}$ models, or at the end of the standard 20M$_{\odot}$ track (which becomes a core collapse supernova as a red supergiant progenitor).

Models for lower-mass stars are also plotted in the lower right panel of Figure \ref{fig:plotFvsCNO}, corresponding to 2M$_{\odot}$, 5M$_{\odot}$, 12M$_{\odot}$, and 15M$_{\odot}$, in order to illustrate that the fluorine abundances in such lower-mass models stay roughly constant, while predicting lower values of $^{16}$O/$^{17}$O with decreasing stellar mass, reaching the lowest values of the oxygen isotopic ratios near $\sim2$M$_{\odot}$. The values for the $^{16}$O/$^{17}$O ratios for the lower-mass Galactic center red giants agree roughly with the trends predicted by the models; the oxygen isotope ratios for the entire sample of red giants presented here will be the subject of a future study.  As in the previously discussed comparisons of elemental abundances, the position of IRS 7 stands out clearly from the other lower-mass Galactic center members with its lower $^{19}$F abundances coupled with a higher $^{16}$O/$^{17}$O ratio.

Before leaving the discussion of comparisons between the observed stellar $^{19}$F abundances and stellar models, we return to the apparently low fluorine abundance in IRS 11 (by $\sim 0.3$ to 0.4 dex) when compared to other NSC red giants with masses less than IRS 7.  The peculiar position of IRS 11 is illustrated in the top panel of Figure \ref{fig:plotlumCF} (A($^{19}$F) versus log g) and the top left panel of Figure \ref{fig:plotFvsCNO} (A($^{19}$F) versus A($^{12}$C): IRS 11 displays a low fluorine abundance given its luminosity and a higher carbon-12 abundance compared to the other NSC luminous cool giants.  In Section \ref{sec:baseline}, the low fluorine abundance was suggested to result from uncertainties in the derived $^{19}$F abundance, as it was derived from only one HF line.  Below, we point to a possible alternative explanation in which IRS 11 has, in fact, undergone fluorine depletion

The three lower-mass NSC stars from Guer\c{c}o et al. \citeyearpar{guerco2022} are IRS 11, BSD 72, and BSD 124, with estimated masses of 4.5M$_{\odot}$, 5.0M$_{\odot}$, and 6.5M$_{\odot}$, respectively. Over this mass range and T$_{\rm eff}$ range of 3600 - 3900 K, core-burning stars can be difficult to distinguish from shell-burning thermally-pulsing asymptotic giant branch (TP-AGB) stars.  Helium-burning thermal pulses drive powerful convection during the TP-AGB phase of stellar evolution, resulting in the dredge-up of both triple-$\alpha$ $^{12}$C and heavy elements produced by slow neutron captures, the s-process: this dredge-up phase is labelled the third dredge-up.  An additional aspect of the third dredge-up in the most massive TP-AGB stars ($\geq 4$M$_{\odot}$) is fluorine destruction via the reaction $^{19}$F($\alpha$,p)$^{22}$Ne (e.g., Palmerini et al. \citeyear{palmerini2019}). Models by Fenner et al. \citeyearpar{fenner2004} that were used and discussed in Smith et al. \citeyearpar{smith2005} predict $^{19}$F depletion in massive AGB stars undergoing hot bottom burning (HBB); models by Legarde et al. \citeyearpar{lagarde2012} also find $^{19}$F destruction in TP-AGB models, but with smaller depletions than those of Fenner et al. \citeyearpar{fenner2004}.  Given the difficulty in determining the exact evolutionary state of red giants near log L$\sim3.8-4.2$L$_{\odot}$ and T$_{\rm eff}\sim3600-3900$ K, core-burning with M$\sim8-9$M$_{\odot}$ or TP-AGB with M$\sim4-7$M$_{\odot}$, it remains possible that IRS 11 is undergoing the third dredge-up or HBB with the resultant destruction of $^{19}$F in the helium-burning regions of thermal pulses. 

\section{Conclusions}

The recent work of  Guer\c{c}o et al. \citeyearpar{guerco2022} derived fluorine abundances in a sample of young (10$^{7}$-10$^{8}$ yr) stellar members of the Nuclear Star Cluster (NSC) located in the very center of the Milky Way, along with one member of the Quintuplet Cluster located some 30 pc from the NSC. The Galactic center stars analyzed in that study were cool (T$_{\rm eff}$ ranging between 3600 K and 3880 K), luminous (log L/L$_{\odot}=3.7$ to 4.9), with masses ranging between $\sim5-15$ M$_{\odot}$. The most luminous star in the NSC, IRS 7 (log L/L$_{\odot}=5.47$), is the focus of the present study. 

The extremely low $^{12}$C and $^{16}$O abundances found by Carr et al. \citeyearpar{carr2000} and Davies et al. \citeyearpar{davies2009} for IRS 7, along with the very large $^{14}$N enhancement derived by Carr et al. \citeyearpar{carr2000}, are confirmed here using spherical MARCS model atmospheres and spherical radiative transfer with the code Turbospectrum (Plez \citeyear{plez2012}). We find [$^{12}$C/Fe]= -0.97, [$^{16}$O/Fe]=-0.55, and [$^{14}$N/Fe]=+1.25, with these values pointing to deep convective mixing in IRS 7.  
New results presented here add the abundance of $^{19}$F, as well as the isotopic ratios of $^{12}$C/$^{13}$C and $^{16}$O/$^{17}$O, to the abundance distribution of IRS 7 and these deep-mixing sensitive abundances were compared to predictions from models of stellar evolution.  

The fluorine abundance obtained for IRS 7 point to a high level of depletion (relative to the baseline F abundance of the NSC) that rivals that observed for carbon, both showing a depletion of $\sim$ 0.8 - 1 dex in IRS 7.  The deep convective mixing of CNO-cycle material, in which $^{19}$F has been destroyed, mixed to the surface of IRS 7 is responsible for the low fluorine abundance.  Such a process had not been observed directly in a massive red supergiant before, so the addition of the $^{19}$F abundance to those of the CNO isotopes can be used to further monitor mixing and refine models of massive stars.  

The abundances of mixing sensitive isotopes and elements in IRS 7 are compared to models of massive stars in order to constrain the evolutionary state of this red supergiant star. Model predictions for the evolution of the fluorine abundances with luminosity are presented for the first time and these were computed in the framework of the stellar models by Ekstr\"om et al. \citeyearpar{Ekstrom2012}. The evolution of fluorine and carbon in the models, along with the results for the C and F abundances for the NSC and IRS 7 (presented in Figures \ref{fig:plotlumCF} and \ref{fig:plotFvsCNO}) show that models predict overall the same behavior as the real stars in the NSC. 
The depleted fluorine abundance in IRS7 illustrates, for the first time, the potential of using the $^{19}$F abundance as a mixing probe in luminous red giants

\section*{Acknowledgements}

\addcontentsline{toc}{section}{Acknowledgements}

{\it Facilities: {Keck Observatory}, {Apache Point Observatory}}.

Software: IRAF (Tody \citeyear{tody1986}, Tody \citeyear{tody1993}), Turbospectrum (Alvarez \& Plez \citeyear{Alvarez_Plez1998}; Plez \citeyear{plez2012}; \href{https://github.com/bertrandplez/Turbospectrum2019}{https://github.com/bertrandplez/Turbospectrum2019}), MOOG (Sneden et al. \citeyear{sneden2012}).

We thank the referee for comments that helped improve the paper. SE and GM have received funding from the European Research Council (ERC) under the European Union's Horizon 2020 research and innovation programme (grant agreement No 833925, project STAREX). CA acknowledges support by the Spanish project PGC2018-095317-B-C21 financed by the MCIN/AEI FEDER “Una manera de hacer Europa”.\\

\subsection*{Data availability}
The data underlying this article are available in the Keck Observatory Archives (KOA), for NIRSpec, and in the SDSS 
Data Release 17 for APOGEE.   These data can be accessed at the following links: \newline
\href{https://nexsci.caltech.edu/archives/koa/}{https://nexsci.caltech.edu/archives/koa/}(NIRSpec) \href{https://www.sdss.org/dr17/irspec/}{https://www.sdss.org/dr17/irspec/}(APOGEE).

\bsp
\label{lastpage}
\end{document}